\documentclass[a4paper, twocolumn]{article}

\usepackage{pifont}
\usepackage{hyperref}
\usepackage{graphicx}
\usepackage{fancyhdr}
\usepackage{amsmath}
\usepackage{lmodern} 
\usepackage{longtable} 
\usepackage{amssymb}
\usepackage{enumitem}
\usepackage{subfigure}
\usepackage{tabularx}
\usepackage{colortbl}
\usepackage{booktabs}
\usepackage{array}
\usepackage{multirow}
\usepackage{url}
\usepackage{xcolor}
\usepackage{float}
\usepackage[utf8]{inputenc}
\usepackage[official]{eurosym}
\usepackage{textcomp}

\usepackage{abstract}
\usepackage{flushend}
\newcolumntype{D}{>{\centering\arraybackslash}X}

\begin{document}

\title{A Framework for Waterfall Pricing Using Simulation-Based Uncertainty Modeling}

\author{
  Nicola Jean\footnote{corresponding author: nicola.jean@gmail.com}, Giacomo Le Pera, Lorenzo Giada, Claudio Nordio\\
  \includegraphics[width = 40mm]{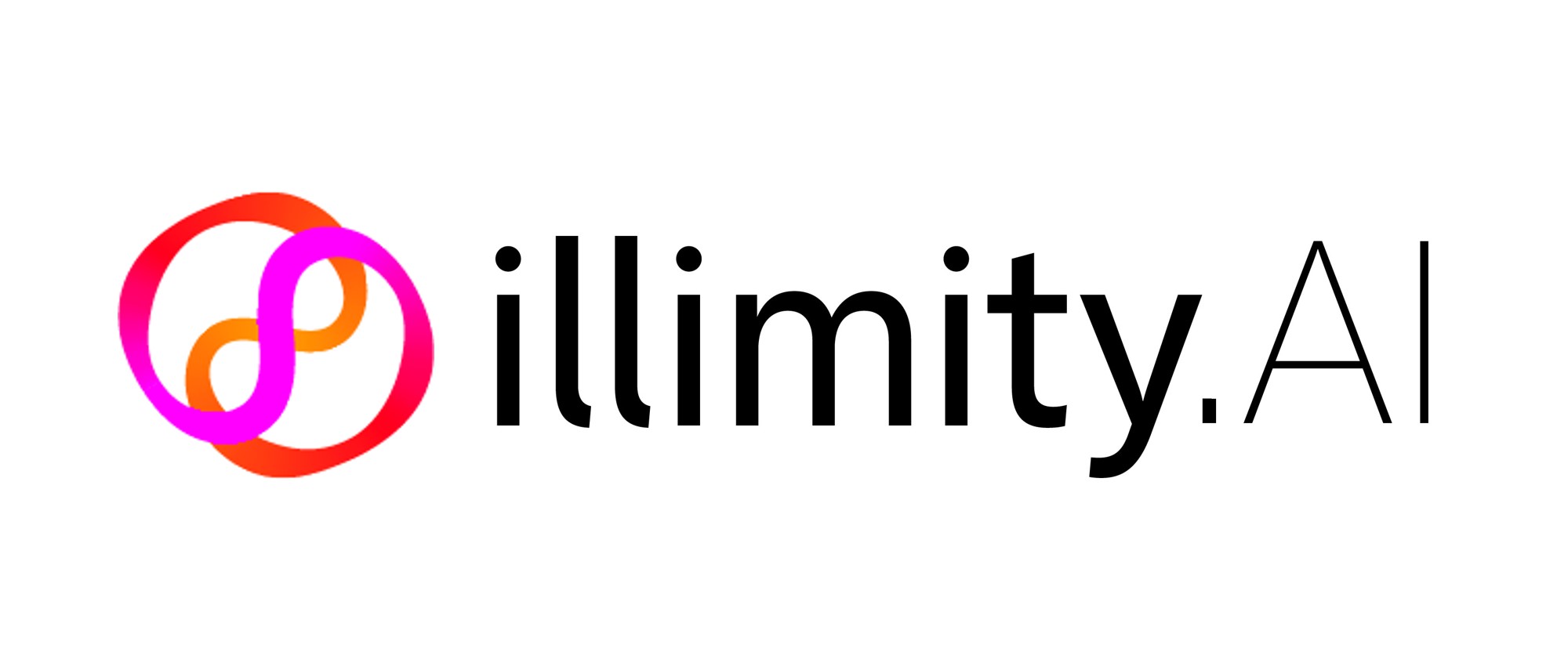}
}
\date{\today \\  [2em] 
{\tt  Working paper\footnote{This paper reflects the authors' opinions and not necessarily those of their employers.}}
}

\twocolumn[
\maketitle
\begin{onecolabstract}
\it We present a novel framework for pricing waterfall structures by simulating the uncertainty of the cashflow generated by the underlying assets in terms of value, time, and confidence levels. Our approach incorporates various probability distributions calibrated on the market price of the tranches at inception. The framework is fully implemented in PyTorch, leveraging its computational efficiency and automatic differentiation capabilities through Adjoint Algorithmic Differentiation (AAD). This enables efficient gradient computation for risk sensitivity analysis and optimization. The proposed methodology provides a flexible and scalable solution for pricing complex structured finance instruments under uncertainty.
\medskip
\end{onecolabstract}
\small{ \bf JEL} Classification codes: C63, G13, G24, C61\\
{ \bf AMS} Classification codes: 91G20, 91G40, 91G60, 91G70
\bigskip
]
\saythanks

\section{Introduction}
Waterfall structures are a fundamental component of structured finance, allocating cash flows from an underlying pool of assets to different tranches based on predefined priority rules. The most explicit example is found in securitization operations, where the credit risk of the assets is distributed among investors in a hierarchical fashion. Pricing these structures requires accurately modeling the uncertainty associated with the underlying assets, particularly in terms of timing and amount of their cash flows, and associated confidence levels, and also accurately encoding all the relevant clauses governing the contract.

Traditional pricing methods often rely on deterministic cash flow projections or simplistic stochastic models. However, these approaches fail to fully capture the complexity of real-world uncertainty and market dynamics. To address these limitations, we propose a simulation-based framework that models uncertainty using a range of probabilistic distributions, calibrated to market-observed tranche prices at inception.

Our implementation is built on PyTorch~\cite{pytorch}, a deep learning framework optimized for tensor computations. PyTorch provides key advantages, including:
\begin{itemize}
    \item Efficient GPU-accelerated computation, enabling large-scale Monte Carlo simulations.
    \item Automatic differentiation through adjoint algorithmic differentiation (AAD), which facilitates efficient gradient calculations for risk management and sensitivity analysis.
    \item Flexibility in defining custom stochastic models, allowing for realistic representation of complex financial instruments.
\end{itemize}

By integrating these capabilities, our framework enhances the accuracy and efficiency of waterfall pricing models.

The rest of the paper is structured as follows: Section 2 discusses the uncertainty modeling approach, Section 3 details the implementation using PyTorch, and Section 4 presents empirical results demonstrating the effectiveness of our methodology. We conclude with insights into potential extensions and applications in structured finance.

\section{Pricing Model and Stochastic Cash Flow Simulation}

The pricing model follows the \textit{No-Arbitrage Theorem}~\cite{harrison1979martingales,harrison1981martingales}, which states that if no arbitrage opportunities exist, then for any tradable asset $A$ and numéraire $B$, there exists at least one equivalent measure $\mathbb{Q}$ under which their ratio is a martingale:

\begin{equation}
    \frac{A(t)}{B(t)} = \mathbb{E}^{\mathbb{Q}}_t\left[\frac{A(T)}{B(T)}\right].
\end{equation}

In a complete market, the risk-neutral pricing measure is unique. However, when dealing with securitizations backed by non-tradable assets, the market is incomplete, and the choice of pricing measure becomes non-unique and necessarily model-dependent.

A natural and practical choice in this context is the \emph{actuarial measure} (also known as the \emph{real-world} or \emph{historical} measure), under which the statistical properties of the underlying assets can be inferred directly from observed historical data. In this framework, the numéraire is typically taken to be the bank account $B(t)$.

Importantly, it is the introduction of a \emph{haircut} parameter to the dynamics of the underlying—typically reflecting credit risk, illiquidity, or conservatism in projection—that causes the pricing measure to deviate from risk-neutrality. The resulting dynamics no longer reflect a martingale under the risk-neutral measure, but instead incorporate risk premiums and structural adjustments aligned with internal valuation policies.

The general pricing formula under the actuarial measure is given by:

\begin{equation}
    A(t) = B(t) \mathbb{E}^{\mathbb{Q}}_t\left[\frac{A(T)}{B(T)}\right].
\end{equation}

For securitization tranches, expected future cash flows $CF_i$ paid at time $t_i$ are discounted using the risk-free curve $D(t,t_i)$. The present value (PV) at time $t$ of all future cash flows is computed as:

\begin{equation}
    PV_t = \sum_{i=1}^{T} D(t,t_i) \cdot \mathbb{E}^{\mathbb{Q}}_t[CF_i] = \sum_{i=1}^{T} \frac{\mathbb{E}^{\mathbb{Q}}_t[CF_i]}{(1 + r_{t_i})^{t_i}},
\end{equation}
where $r_{t_i}$ is the risk-free discount rate from $t$ to $t_i$. When amortization is present, the cash flows $CF_i$ include both interest payments $I_{t_i} = N_{{t_i-1}} \cdot s_{t_i}$ and principal repayments $P^i_{t_i} = N_{{t_i-1}} - N_{t_i}$, where $N_{t_i}$ is the remaining principal after the payment in $t_i$ and $s_{t_i}$ is the interest paid on period from $t_i$ to $t_{i+1}$. In the previous formula we assume no dynamics on the interest rates component and this allows us to move $ D(t,t_i) $ out of the expectation formula.

\subsection{Stochastic Cash Flow Simulation}

Unlike the common practice in market risk, where risk drivers are often defined under the assumption that historical series of the underlying are independent and identically distributed (i.i.d.)~\cite{meucci2005risk}, our methodology does not rely on this assumption. Instead, we develop a general framework for modeling cash flows that remains applicable even in highly illiquid markets on non-tradable assets where the martingale property must be postulated based on general principles rather than empirical data.

To model uncertainty in (net) cash flows available to the waterfall, we introduce three stochastic engines, that are applied at the aggregate level. Thus the present framework is not bottom-up (or microscopic), in the sense that there is no assumption on the dynamics of the individual underlying assets.

1. \textbf{OneSigmaCashFlowEngine}: This engine models the logarithm of cash flows using a normal distribution incorporating a drift term $\mu$, accounting for systematic trends in the cash flow evolution:

   \begin{equation}
       \log {CF_i} \sim   \mathcal{N} \left( \log\overline{CF_{i,t_i}} + (\mu - \frac{\sigma^2}{2})t_i,
        \sigma^2 t_i \right).
   \end{equation}
The drift $\mu$ allows for long-term deviations from the initial business plan expectation, while the volatility $\sigma$ captures short-term fluctuations. In our setup we usually set $\mu$ equal to 0, since we noticed that a similar effect is more realistically captured by the third Engine below.

2. \textbf{MultipleStochasticTime}: With this engine, a fraction $p$ of each cash flow $CF_i$ is received at a different arrival time $t_i$, where the interarrival times $\tau_i = t_i - t_{i-1}$ follow a Pareto distribution with exponent $\alpha>0$ and expected value 1, with correlation $\rho$ between any pair of interarrival times \cite{brigo2010liquidity}:

\begin{align}
    \tau_i &\sim \mathrm{Pareto}(\alpha) \notag \\
    \mathbb{E}[\tau_i] &= 1 \notag \\
    \mathrm{Corr}(\tau_i, \tau_j) &= \rho, \quad i \neq j
\end{align}

   The Pareto distribution has shape parameter $\alpha$ and scale parameter $x_m = \frac{\alpha-1}{\alpha}$, with probability density function:
   \begin{equation}
       f(x) = \frac{\alpha x_m^{\alpha}}{x^{\alpha+1}} \quad \text{for } x \geq x_m
   \end{equation}

   The arrival times are defined as the cumulative sum of interarrival times: $t_i = \sum_{k=1}^{i} \tau_k$. The correlation between any two arrival times $t_i$ and $t_j$ (assuming $i < j$) can be derived as follows:
   \begin{align}
   \mathrm{Cov}(t_i, t_j) &= \mathrm{Cov}\left(\sum_{k=1}^{i} \tau_k, \sum_{l=1}^{j} \tau_l\right) \nonumber \\
   &= i\sigma_\tau^2[1 + \rho(j-1)]
   \end{align}

   The variances of arrival times are $\mathrm{Var}(t_i) = i\sigma_\tau^2[1 + \rho(i-1)]$ and $\mathrm{Var}(t_j) = j\sigma_\tau^2[1 + \rho(j-1)]$. Therefore, the correlation between arrival times is:
   \begin{equation}
   \mathrm{Corr}(t_i, t_j) = \sqrt{\frac{i}{j}} \cdot \frac{\sqrt{1 + \rho(j-1)}}{\sqrt{1 + \rho(i-1)}}
   \end{equation}

   This formula has the properties that $\mathrm{Corr}(t_i, t_j) = \sqrt{\frac{i}{j}}$ when $\rho = 0$ (independent interarrival times) and $\mathrm{Corr}(t_i, t_j) = 1$ when $\rho = 1$ (perfectly correlated interarrival times).

3. \textbf{SpreadCashFlowEngine}: This engine adjusts cash flows based on an aggregated adjustment factor $w$:
   \begin{equation}
   \hat{CF}_{i} =  w \cdot CF_{i}.
   \end{equation}

Collectively, these engines capture volatility, heavy-tailed distributions, temporal correlation, and spread adjustments (haircut) of the cash flows, providing a dynamic and robust modeling approach for the valuation of securitization tranches. We have found these dynamics sufficient to calibrate the model at inception, where the prices of the various tranches are market implied. Our calibration process leverages scipy~\cite{scipy} genetic algorithm implemented in the differential evolution module. The optimization function uses the maximum error as parameter to define the convergence level.

These engines are applied sequentially to model different sources of uncertainty:

\begin{itemize}
    \item \textbf{OneSigmaCashFlowEngine:} models uncertainty in the \emph{amount} of each cash flow at every time step.
    \item \textbf{MultipleStochasticTime:} captures uncertainty in the \emph{timing} of the cash flows.
    \item \textbf{SpreadCashFlowEngine:} reflects uncertainty in the \emph{baseline forecast}, affecting the overall evaluation of future cash flows.
\end{itemize}

We have introduced a set of up to 6 parameters ($\sigma$, $\mu$, $p$, $\alpha$, $\rho$, $w$) that can to be calibrated in order to reproduce market prices of the tranches, if available, or might be extracted from historical data for the underlying asset class. In any case, even if the model is rather parsimonious, a certain degree of overparametrization is unavoidable in order to obtain a more realistic behavior of the collections. Thus the robustness of the results has to be tested against small changes in the objective function.

\subsection{Algorithmic Adjoint Differentiation}
Securitizations are complex financial instruments characterized by highly nonlinear and recursive cash flow structures. These instruments are designed to create a \textit{payment waterfall}, where risk is distributed asymmetrically among different tranches. Senior tranches benefit from higher protection, while Mezzanine and junior tranches bear increased marginal risk.

To achieve this structured payout, various nonlinear mechanisms such as triggers, subordination, and loss absorption are embedded within the transaction, making securitizations highly exotic derivatives. Due to this complexity, a Monte Carlo simulation approach is typically required for pricing, as closed-form solutions are infeasible in most cases.

Computing risk sensitivities (i.e., Greeks) in these structures using traditional finite difference methods is computationally expensive. Each perturbation of a risk factor necessitates a full revaluation of the entire securitization structure, leading to significant runtime overhead. Given the recursive nature of cash flows and the dependence on simulated scenarios, this brute-force approach quickly becomes impractical.

To efficiently compute sensitivities, \textbf{Algorithmic Adjoint Differentiation (AAD)}~\cite{AAD_ref} offers a powerful alternative. AAD allows for the computation of gradients with respect to multiple risk factors in a single backward pass, dramatically reducing computational cost.

In Python, \texttt{PyTorch} provides an efficient computational framework for implementing AAD due to its native support for automatic differentiation. By defining tensors with \texttt{requires\_grad=True}, PyTorch automatically tracks operations and computes gradients via reverse-mode differentiation. This enables efficient gradient computation, even for complex securitization models.

However, certain best practices and caveats should be considered when using PyTorch for AAD in structured finance applications:

\begin{itemize}
    \item \textbf{Avoid division by small numbers:} Operations involving division can introduce NaNs in the computational graph, breaking the gradient flow. Regularization techniques or adding small $\epsilon$ values can mitigate this issue.

    \item \textbf{Do not use hard clipping or casting:} Instead of explicit thresholding operations (e.g., \texttt{torch.clamp()} or casting to integers usually done to work on vector indexes), use smooth approximations such as the \textbf{double sigmoid mask function}:
    \begin{equation}
      \text{mask} = \sigma_k (\tau - i) \cdot \left( 1 - \sigma_k (\tau - (i+1)) \right),
    \end{equation}
    where $\sigma_k(x)$ is the sigmoid function:
    \begin{equation}
      \sigma_k(x) = \frac{1}{1 + e^{-kx}}.
    \end{equation}
    This formulation ensures a smooth and differentiable transition. We adopted this approach to prevent gradient vanishing when simulating the collection time of future cash flows.

    \item \textbf{Minimize explicit loops:} PyTorch operates optimally with vectorized computations. Avoid explicit \texttt{for} loops and leverage tensor operations to maximize GPU efficiency.

    \item \textbf{Avoid conditional branching (if-statements):} Control flow statements disrupt the tensor computation graph, limiting differentiation capabilities. Whenever possible, use smooth approximations such as soft indicators or differentiable threshold functions.

    \item \textbf{Ensure consistent random number generation:} For reproducibility in Monte Carlo simulations, ensure that the seed is explicitly set.

    \item \textbf{Be cautious with ReLU and max operations:} Functions like \texttt{ReLU} and \texttt{torch.max()} truncate gradient propagation beyond the first derivative, which may lead to unstable sensitivity calculations. Alternative smooth approximations such as Softplus:
    \begin{equation}
      \text{Softplus}(x) = \frac{1}{\beta} \log(1 + e^{\beta x})
    \end{equation}
    can help maintain gradient flow.

\end{itemize}

\subsection{A simple application}

To demonstrate why securitizations need to be priced taking into account the stochastic dynamical nature of the underlying we have built a toy model based on recovery plans from a pool of assets\footnote{\frenchspacing this situation arises e.g. in securitizations of non-performing loans, or of real estate assets}. The semi-annual cash flows include asset rental income, collection fees, and stochastic asset sales times exponentially distributed.

The key components are:

\begin{itemize}
    \item \textbf{Asset Types:} Each asset type $i$ has an initial notional value $V_i$, a rent yield $r$, a depreciation factor $\delta_i$, and an exponential parameter $\lambda_i$ governing the time to sale.

    \item \textbf{Rental Income:} Rental income is accrued semi-annually as
    \[
    \text{NetRent}_{i} = \frac{r \cdot V_i}{2} \cdot (1 - \text{fee}),
    \]
    where \texttt{fee} is a fixed rent collection fee.

    \item \textbf{Sale Timing:} The time of sale for each asset is sampled from a \emph{Gaussian copula} to introduce correlation across asset types. Specifically:

    \item \textbf{Sale Proceeds:} Assets are sold at their depreciated value:
    \[
    \text{SalePrice}_i = V_i \cdot \delta_i^{t_i},
    \]
    where $t_i$ is the number of years until the sale. A random half-period offset is introduced to allow sale times at mid-year points.

    \item \textbf{Cash Flow Aggregation:} For each asset, we add the net rental income at every 6-month period until the sale, and the full sale proceeds at the time of sale. These are aggregated across all assets.
\end{itemize}

Formally, let \( V_0 \) be the initial value of an asset, and let \( T_s \) be the randomly drawn sale time (in years). Let \( h \in \left\{ 0, \frac{1}{2} \right\} \) be a random half-period offset, so that the effective sale time is \( \tilde{T}_s = T_s + h \). Then, the cash flows \( CF_t \) at each 6-month period \( t \) are given by:

\begin{equation}
CF_t = \begin{cases}
\frac{r V_0}{2} (1 - \beta), & t < 2\tilde{T}_s \quad \text{(net rental income)} \\
V_0 \cdot \delta^{\tilde{T}_s}, & t = 2\tilde{T}_s \quad \text{(sale proceeds)} \\
0, & t > 2\tilde{T}_s
\end{cases}
\end{equation}

where \( \beta \) represents collection costs and other fees, and \( \delta \) is the annual depreciation factor.

The total portfolio cash flow is computed by summing the individual asset cash flows across all asset types. Each asset generates periodic rental income until a stochastic sale event occurs—modeled via an exponentially distributed holding time with possible half-period granularity. This setup mimics the kind of irregular but structured cash flow behavior observed in asset pools used in securitizations.

\begin{table*}[!ht]
    \centering
    \begin{tabular}{|c|l|}
        \hline
        \textbf{Parameter} & \textbf{Value} \\
        \hline
        $N_a$ (assets per type) & 20  \\
        $V_0$ (initial asset values) & [1, 1.5, 2, 2.5, 3] (in ~\euro{}~mn) \\
        $r$ (annual rental yield) & 5\%  \\
        $T$ (years) & 10 \\
        $\lambda$ (exp. sale rate per type) & [0.5, 0.4, 0.45, 0.35, 0.3]  \\
        $\delta$ (annual price decay factors) & [0.98, 0.97, 0.99, 0.96, 0.95]  \\
        $\beta$ (collection fee) & 10\% \\
        $\rho$ (correlation between asset types) & 0.5  \\
        \hline
    \end{tabular}
    \caption{Default parameters for the simulated asset cash flows}
    \label{tab:cashflow_params}
\end{table*}

On top of this pool we have built a waterfall with Senior, Mezzanine and Junior tranche. A limited recourse loan tranche is added as well\footnote{this is a common mechanism to create a cash reserve}. Mezzanine payments are subject to different regimes according to cumulative collection rate (CCR) conditions\footnote{total collections are compared to an initial contractual profile}.

The structured cash flow waterfall dictates the sequential allocation of available cash flows at each time step. The prioritization is as follows:

\begin{enumerate}
\item \textbf{Senior Expenses and Reserves:} Payment of expenses related to the Senior tranche, reserves for Master SPVs, and reserves for Master LeasesCo.
\item \textbf{Servicer Fees:} Fees due to the Master Servicer and Special Servicer.
\item \textbf{LRL Interest Payments:} Interest payments on the Limited Recourse Loan (LRL).
\item \textbf{Senior Note Interest Payments:} Interest due on the Senior tranche, linked to the EURIBOR floating rate.
\item \textbf{Mezzanine Interest Payments:} Interest on the Mezzanine tranche, subject to the CCR threshold. If the CCR falls below a predefined threshold, payments will be postponed after the Senior Notional Payment.
\item \textbf{Cash Reserve Allocation:} A reserve amount is allocated based on a target rate applied to the Senior tranche's notional balance.
\item \textbf{LRL Notional Payments:} Repayment of the LRL notional balance by an amount liked to the residual Senior tranche notional balance.
\item \textbf{Senior Notional Payments:} Repayment of the Senior tranche notional balance.
\item \textbf{Mezzanine Deferred and Post-Enforcement Interest:} Pending and post-enforcement interest payments for the Mezzanine tranche, dependent on the CCR level.
\item \textbf{Mezzanine Notional Payments:} Principal repayment of the Mezzanine tranche.
\item \textbf{Junior Spread Payments:} Distribution of any remaining funds to junior tranche holders as spread income.
\item \textbf{Junior Notional Payments:} Repayment of the junior tranche notional balance.
\item \textbf{Junior Variable Returns:} Any final surplus cash flows allocated to the junior tranche as variable returns.
\end{enumerate}

The parameters used in the waterfall definition are defined in Table~\ref{tab:tranche_params}.

\begin{table}[h]
    \centering
    \begin{tabular}{|l|l|}
        \hline
        \textbf{Component} & \textbf{Value} \\
        \hline
        Senior Note Interest & Euribor $+~2.5\%$ \\
        Mezzanine Note Interest & Euribor $+~5\%$ \\
        Junior Note Interest & $10\%$ \\
        LRL Interest & Euribor $+~0.2\%$ \\
        \hline
        Senior Notional & 135 ~\euro{}~mn \\
        Mezzanine Notional & 31.5 ~\euro{}~mn\\
        Junior Notional & 13.5 ~\euro{}~mn \\
        LRL Notional & 5.805 ~\euro{}~mn \\
        Total CF base scenario & 206 ~\euro{}~mn \\
        CCR Threshold & $0.9$ \\
        \hline
    \end{tabular}
    \caption{Tranche structure and parameters}
    \label{tab:tranche_params}
\end{table}

In this setup if we set Senior Price 100.0\% notional balance, Mezzanine Price 30.0 \% notional balance and Junior Price 5.0 \% notional balance,  we get optimal parameters (max calibration error of 2 basis points):

\begin{table}[h]
    \centering
    \begin{tabular}{|c|c|}
        \hline
        \textbf{Parameter} & \textbf{Value} \\
        \hline
        $\alpha$ & 4.6305 \\
        $\rho$ & 0.5 \\
        $p$ & 0.8646 \\
        $\sigma$ & 0.1053 \\
        $w$ & 0.7571 \\
        \hline
    \end{tabular}
    \caption{Parameters for OneSigmaMultipleStochasticTime}
    \label{tab:one_sigma_params}
\end{table}

In securitizations, it is standard practice to express tranche prices as a percentage of their initial notional balance. We will adopt this convention throughout the remainder of this work.

\begin{figure}[h]
    \centering
    \includegraphics[width=0.8\linewidth]{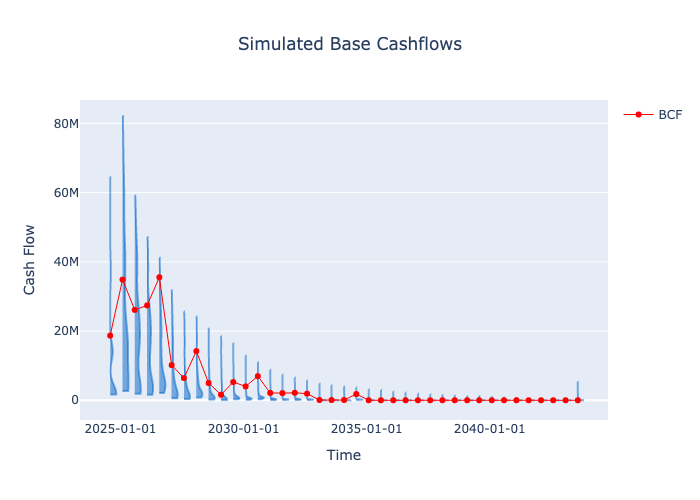}
    \caption{Craviola Plot}
    \label{fig:toy_craviola}
\end{figure}

\begin{figure}[h]
    \centering
    \includegraphics[width=0.8\linewidth]{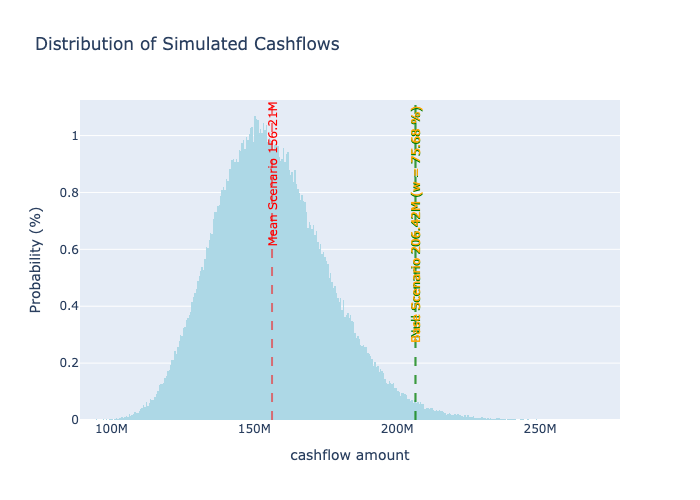}
    \caption{Total Cash Flow}
    \label{fig:toy_meannull}
\end{figure}

The generated dynamics is depicted time by time in Figure~\ref{fig:toy_craviola}, and collapsed on the start date in Figure~\ref{fig:toy_meannull}

\begin{figure}[h]
    \centering
    \includegraphics[width=0.8\linewidth]{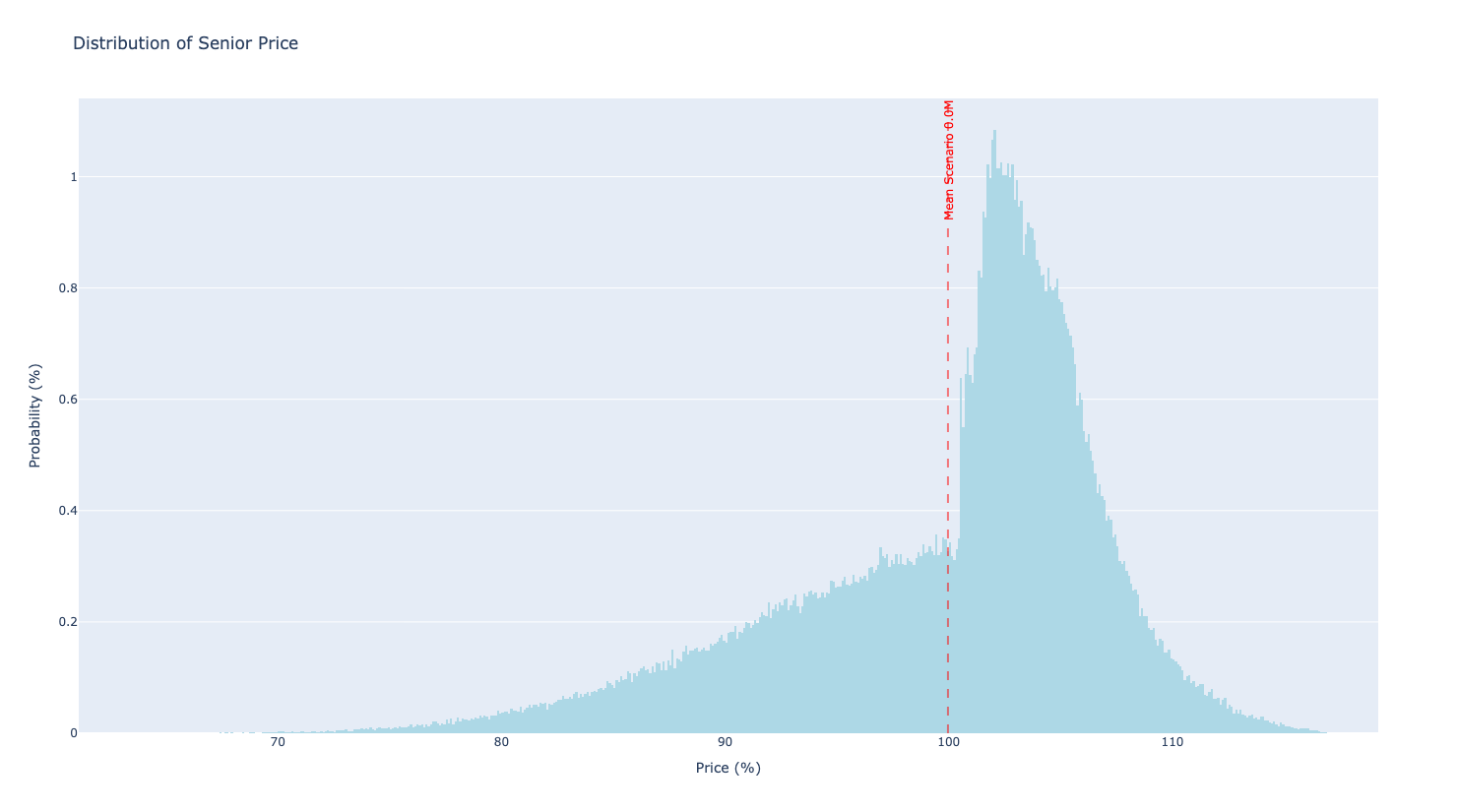} 
    \caption{Senior price distribution}
    \label{fig:price_senior}
\end{figure}

\begin{figure}[h]
    \centering
    \includegraphics[width=0.8\linewidth]{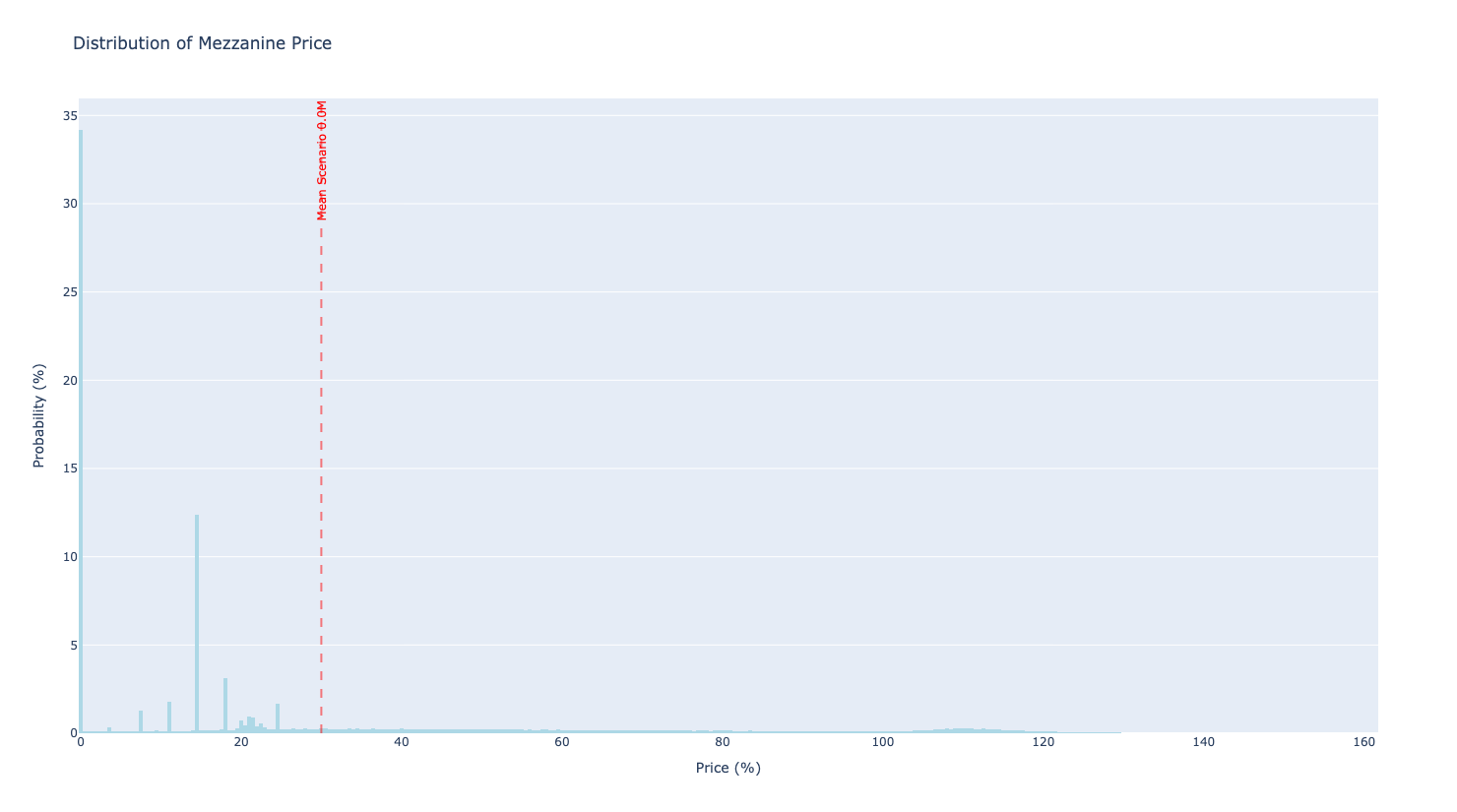}
    \caption{Mezzanine price distribution}
    \label{fig:price_mezzanine}
\end{figure}

The dynamics described above generate a distribution of prices for the Senior and Mezzanine tranches, as shown in Figure~\ref{fig:price_senior} and Figure~\ref{fig:price_mezzanine}. A simple model without stochastic elements—typically used to analyze such complex structures—would fail to capture the insights provided by the full price dynamics. For instance, the Mezzanine tranche is particularly challenging to evaluate and manage. A basic analysis based on average prices would overlook key irregularities caused by the CCR-triggered inversion of payment priorities.

To identify the weak points of the structure, we leverage Algorithmic Adjoint Differentiation. Analyzing the resulting sensitivities provides insight into the factors to which the structure is most vulnerable.

\begin{table}[h]
    \centering
    \renewcommand{\arraystretch}{1.2} 
    \setlength{\tabcolsep}{6pt} 
    \begin{tabular}{|l|r|}
        \hline
        \textbf{Parameter} & \textbf{Gradient / Value} \\
        \hline
        $\alpha$ & 0.052083 \\
        $p$ & -0.909420 \\
        $\sigma$ (param) & -46.772629 \\
        $w$ & 64.733177 \\
        DV01 (up 1 bp) & 0.009848 \\
        BV01 (up 1 euro) & 6.869932 \\
        \hline
    \end{tabular}
    \caption{Gradients and Sensitivities of the Senior Tranche in terms of a 100.0 price. }
    \label{tab:gradients_senior}
\end{table}

BV01 stands for the sensitivity to underlying cash flow changes. The Senior price of our structure is positively affected both by increases in the underlying cash flows and by higher interest rates.
\begin{table}[h]
    \centering
    \renewcommand{\arraystretch}{1.2} 
    \setlength{\tabcolsep}{6pt} 
    \begin{tabular}{|l|r|}
        \hline
        \textbf{Parameter} & \textbf{Gradient / Value} \\
        \hline
        $\alpha$ & -0.226221 \\
        $p$ & -1.233831 \\
        $\sigma$ (param) & 84.324936 \\
        $w$ & 327.795898 \\
        DV01 (up 1 bp) & -0.020087 \\
        BV01 (up 1 euro) & 38.900495 \\
        \hline
    \end{tabular}
    \caption{Gradients and Sensitivities of the Mezzanine Tranche in terms of a 100.0 price.}
    \label{tab:gradients}
\end{table}

The Mezzanine price is positively impacted by increases in the underlying cash flows but negatively affected by higher interest rates. When adding an interest rate hedge to the structure, it is important to consider the different interest rate sensitivities.

\begin{figure*}[h]
    \centering
    \includegraphics[width=0.8\textwidth]{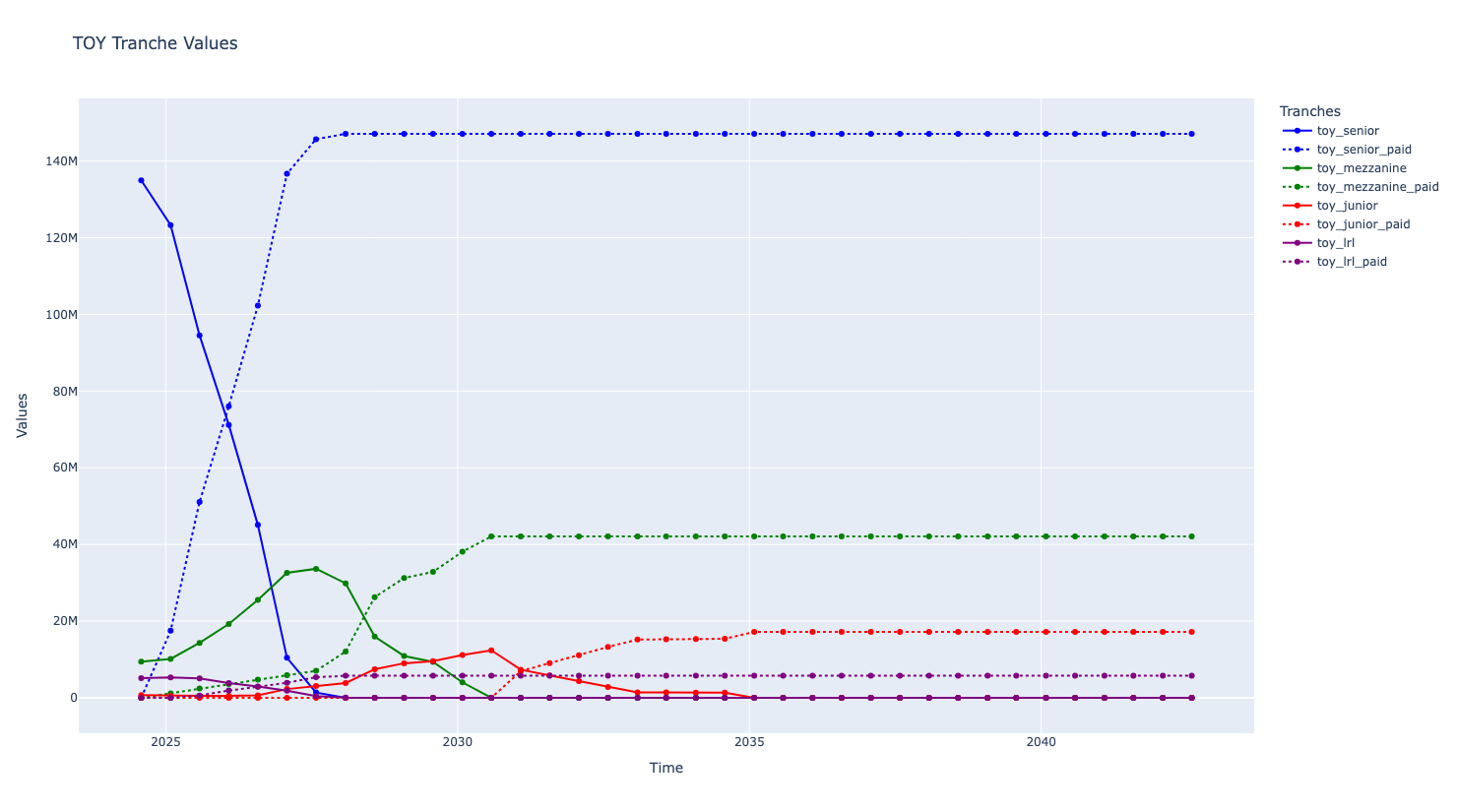}
    \caption{Time Value of the Tranches}
    \label{fig:time_lapse}
\end{figure*}

Our framework allows also to generate forward prices by simply moving the evaluation date forward in time. For this simple toy model case we get Figure~\ref{fig:time_lapse} .

The previous chart represents the expected price profile time by time and gives information on the stability of the structure throughout time at inception market forecast.

With all tranche prices now properly evaluated using the stochastic model, we can translate the results into the standard intrinsic valuation framework based on a single scenario. This involves computing the Internal Rate of Return (IRR), Z-spread, annuity equivalent, and asset swap spread for each tranche. These quantities allow us to translate a complex model into a simpler representation, effectively abstracting the underlying complexity and making the system more interpretable.

\newpage

\begin{table*}[!ht]
    \centering
    \begin{tabular}{|l|c|c|c|c|}
        \hline
        \textbf{Tranche} & \textbf{Annuity} & \textbf{IRR (\%)} & \textbf{Z-spread (\%)} & \textbf{ASW (\%)} \\
        \hline
        Senior     & 154.485   & 0.0555 & 0.0232 & 0.0234 \\
        Mezzanine  & 417.5819  & 0.4517 & 0.3479 & 0.2166 \\
        Junior     & 1262.4413 & 0.5815 & 0.4342 & 0.0832 \\
        LRL        & 202.1887  & 0.0654 & 0.0348 & 0.0324 \\
        \hline
    \end{tabular}
    \caption{Intrinsic valuation metrics for each tranche in the toy example}
    \label{tab:toy_tranche_metrics}
\end{table*}

The quantities reported in Table~\ref{tab:toy_tranche_metrics} have been computed using the following formulae:

\begin{itemize}
    \item \textbf{IRR}:
    \begin{equation}
        \text{IRR} = r^* \quad \text{such that} \quad \sum_{i=1}^{N} \frac{C_i}{(1 + r^*)^{t_i}} = P
    \end{equation}
    where:
    \begin{itemize}
        \item \( C_i \) - Cashflows at time \( t_i \).
        \item \( t_i \) - Year fractions associated with each cashflow.
        \item \( P \) - Observed price.
        \item \( r^* \) - Internal Rate of Return (IRR), solved numerically.
    \end{itemize}

    \item \textbf{Z-spread}:
    \begin{equation}
        P = \sum_{i=1}^{N} C_i \cdot D_i \cdot e^{-t_i \cdot z}
    \end{equation}
    where:
    \begin{itemize}
        \item \( C_i \) - Cashflows at time \( t_i \).
        \item \( D_i \) - Discount factors.
        \item \( t_i \) - Year fractions associated with each cashflow.
        \item \( z \) - Z-spread, solved numerically.
        \item \( P \) - Observed price.
    \end{itemize}

    \item \textbf{Annuity}:
    \begin{equation}
        A = \frac{100}{\max(N - L, \epsilon)} \sum_{i=0}^{N-T} D_{i+T} \cdot (A_i - L) \cdot Y_{i+T}
    \end{equation}
    where:
    \begin{itemize}
        \item \( N \) - Actual notional balance.
        \item \( L \) - Last amount.
        \item \( A_i \) - Amortizing plan.
        \item \( D_i \) - Discount factors.
        \item \( Y_i \) - Year fractions.
        \item \( T \) - Time lapse.
        \item \( \epsilon \) - Small constant to prevent division by zero.
    \end{itemize}

    \item \textbf{ASW}:
    \begin{equation}
        \text{ASW} = \frac{P_0 - P}{A}
    \end{equation}
    where:
    \begin{itemize}
        \item \( P_0 \) - Price in the null scenario.
        \item \( P \) - Observed price.
        \item \( A \) - Annuity.
        \item \text{ASW} - Asset Swap Spread.
    \end{itemize}
\end{itemize}

\section{Conclusions}\label{sec:conclusions}
We have developed a framework for evaluating securitizations of complex assets, incorporating simulations that capture their dynamics in terms of value, time, and total value uncertainty. This framework not only generates prices but also provides sensitivities, time value estimates, and other key metrics essential for risk management and structural robustness. Future enhancements will focus on integrating interest rate derivatives, enabling "what-if" analysis, and handling securitizations of securitizations. Additionally, we aim to experiment with more granular dynamics to further stress-test the stability of results.

\section*{Acknowledgments}

We would like to extend our gratitude to Fabio Gallotti for the implementation of all the needed APIs and backend services that made it possible to execute the evaluation process in a cloud-native setup.

\bibliographystyle{plain}
\bibliography{biblio}

\end{document}